\documentclass[]{emulateapj}
\usepackage{amsmath}
\usepackage{amsfonts}
\usepackage{amssymb}

\def\degree{\ifmmode {^\circ}\else {$^\circ$}\fi}
\def\kms{\ifmmode {\rm km~s^{-1}}\else $\rm km~s^{-1}$\fi}
\shorttitle{Identifying Star Streams in the Halo}
\begin{document}

\slugcomment{ApJ submitted}

\title{Identifying Star Streams in the Milky Way Halo}

\author{Charles King III} % \altaffilmark{1}}
\affil
{Smithsonian Astrophysical Observatory, 60 Garden Street, Cambridge, MA 02138\\
Pleiades Consulting Group Inc., Post Office Box 531, Lincoln, MA 01773}
\email{cking@cfa.harvard.edu}

\author{Warren R.\ Brown}
\affil
{Smithsonian Astrophysical Observatory,
60 Garden Street,
Cambridge, MA 02138}
\email{wbrown@cfa.harvard.edu}

\author{Margaret J.\ Geller}
\affil
{Smithsonian Astrophysical Observatory,
60 Garden Street,
Cambridge, MA 02138}
%\email{mgeller@cfa.harvard.edu}

\and

\author{Scott J.\ Kenyon}
\affil
{Smithsonian Astrophysical Observatory,
60 Garden Street,
Cambridge, MA 02138}
%\email{skenyon@cfa.harvard.edu}

%\altaffiltext{1}{Pleiades Consulting Group Inc., Post Office Box 531, Lincoln, MA 01773}

\begin{abstract}

We develop statistical methods for identifying star streams in the halo of
the Milky Way galaxy that exploit observed spatial and radial velocity
distributions. Within a great circle, departures of the observed spatial
distribution from random provide a measure of the likelihood of a potential
star stream. Comparisons between the radial velocity distribution within a
great circle and the radial velocity distribution of the entire sample also
measure the statistical significance of potential streams. The radial
velocities enable construction of a more powerful joint statistical test for
identifying star streams in the Milky Way halo. Applying our method to halo
stars in the Hypervelocity Star (HVS) survey, we detect the Sagittarius
stream at high significance. Great circle counts and comparisons with
theoretical models suggest that the Sagittarius stream comprises 10\% to
17\% of the halo stars in the HVS sample. The population of blue stragglers
and blue horizontal branch stars varies along the stream and is a potential
probe of the distribution of stellar populations in the Sagittarius dwarf
galaxy prior to disruption.

\end{abstract}


\section{INTRODUCTION}
\label{sec: Intro}

        The timescale for halo stars to exchange their energy and angular
momentum is long compared to the age of the Milky Way galaxy. Thus, the
stellar halo of the Milky Way provides a unique laboratory for studying
hierarchical galaxy formation. Theoretical simulations show that the
remnant debris of hierarchical galaxy formation should be visible as star
streams in the Milky Way halo \citep{Johnston96, Harding01, Abadi03b,
Bullock05, Font06}. Indeed, observers have discovered the tidal debris of
the Sagittarius dwarf galaxy \citep{Ibata94} wrapping around the Milky Way
\citep{Majewski03} and other stellar overdensities throughout the stellar
halo \citep[e.g.,][]{Belokurov06}. The vast majority of these detections
come from star counts without radial velocity measurements.

        We describe a powerful technique for identifying tidal star
streams that exploits the additional information provided by a complete
radial velocity survey. With large stellar radial velocity surveys such
as RAVE \citep{Steinmetz06, Zwitter08} and SEGUE \citep{Yanny09} now
available, approaches including radial velocities enable better
constraints on structure in the halo.

To identify statistically significant tidal structures, we examine the 
surface number distribution and the radial velocity distribution of stars
in great circles. Both the surface density and the radial velocity distributions 
probe the origin of structure in the halo. A radial velocity test for 
structure applies independently of the underlying spatial distribution. 
Our technique opens up the possibility of finding tidal star streams based 
on structure in velocity space. Combining our spatial and radial velocity 
statistical tests provides a more powerful joint test for identifying star 
streams.

Previous studies of positions and velocities have focused on known tidal
streams, such as Sagittarius \citep[e.g.,][]{Yanny09, Ruhland11}, or have
analyzed disconnected patches scattered across the sky
\citep{Starkenburg09, Schlaufman09}.  Known tidal debris, however, is
contiguous over areas of hundreds or thousands of square degrees
\citep{Johnston99,Ibata01}.

As a first demonstration of our technique, we search for structure in the
HVS radial velocity survey, a complete, non-kinematically selected sample
of late B-type stars covering a contiguous area of more than 8400
deg$^{2}$ on the sky \citep{brown05, brown06b, brown09a, brown10}. The HVS
survey targets blue stragglers and blue horizontal branch (BHB) stars
within the halo and is distinct in its uniform velocity data for a
well-defined sample of potential halo stars.

We detect significant structure in the sample of halo stars within the
HVS survey.  Most of this structure is attributed to the Sagittarius
stream. We compare our detections with the \citet{Law10a} {\it N}-body
model for the Sagittarius dwarf, derive estimates for the fraction of
HVS stars within the Sagittarius stream, and assess the relative
fractions of blue stragglers and BHB stars along the stream.

In \S \ref{sec:IDStreams}, we describe the statistical technique. In
\S \ref{sec:DataSection}, we apply this technique to the HVS survey.
In \S \ref{sec:SgrBlueStars}, we compare the survey stars with the
$N$-body model of \citet{Law10a} in order to explore the nature and surface
density of stars within the Sagittarius stream. We conclude in \S
\ref{sec:Conclusion}.

% Figure 1
\begin{figure} \begin{center} \includegraphics[width=3.25in]{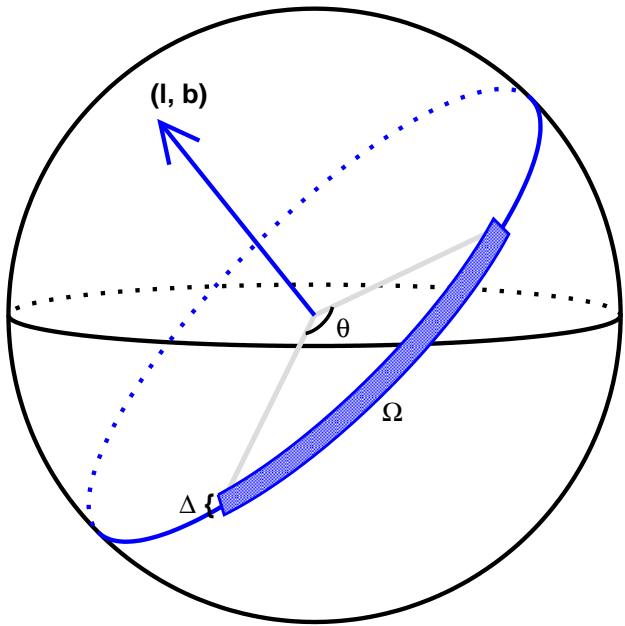} 
\caption{
	Solid angle, $\Omega $, subtended by a great circle segment with arc length $\protect\theta $, width $\Delta $, and pole at galactic 
coordinates $(l,b)$. 
\label{fig: graphic} } \end{center} 
\end{figure}

\section{TECHNIQUE FOR IDENTIFYING STAR STREAMS}\label{sec:IDStreams}

Identifying structures in the Galactic halo is a challenging statistical
problem. The structures are low contrast and may extend across the entire sky.
For most stars in the Galactic halo, we have ready access to only half of phase
space. Radial velocity and angular position are the only robust coordinates
available for large samples of stars more distant than $\sim 10$ kpc
\citep[\emph{e.g.,}][]{Schlaufman09}.

Methods of identifying star streams and other substructures in the Milky
Way halo include pole counts \citep{Lynden-Bell95}, great circle counts
\citep{Johnston96}, star count maps \citep[e.g.,][]{Ibata02b,
Belokurov06}, group finding algorithms \citep{Sharma11, Xue11}, and
combined analyses of the spatial and radial velocity distributions of halo
populations \citep[\emph{e.g.,}][]{Schlaufman09}. Most of these methods
include an assessment of the statistical significance of the structures
they detect.

The method we develop contains elements of previous approaches refined for
application to radial velocity samples that are complete over some
substantial portion of the sky. Our approach has three steps: (1) we
analyze great circle counts based on a nearly uniform, dense grid of
sampling poles, (2)  we compare the great circle radial velocity
distribution with the radial velocity distribution for the entire survey,
and (3) we combine the great circle counts and velocity distribution
comparison into a single statistic. We estimate the significance level, or
probability of false positives, among the structures we detect.

\subsection{Great Circle Counts \label{sec: Counts}}

The orbits of tidal debris from disrupted satellite galaxies in the Milky Way
halo persist for several Gyr. Because the Galactic halo is approximately
spherical, a satellite's orbit in the sky should approximate a great circle.
The ``fossil'' debris should be distributed along the orbit of the satellite
\citep[\emph{e.g.,}][]{Johnston96,Bullock05}.

\cite{Johnston96} propose using star counts in great circles for
detecting tidal debris against background halo stars. They define a
great circle by its width and by the position of its pole relative to
the Galactic pole. The pole is the unit normal to the plane of the
great circle. The latitude and longitude of the pole on the unit sphere
then determine the location of its corresponding great circle, as
illustrated in Figure \ref{fig: graphic}. \cite{Johnston96} compare
counts of stars observed within a great circle with the number of stars
expected from a random distribution of stars on the sky. We generalize
this approach for an observational sample that covers only a portion of
the sky. For simplicity, we refer to the number of stars in a great
circle segment as the great circle count.

For a great circle segment of arc length $\theta $ and width $\Delta $
within the sample region, the solid angle subtended the segment is$\
\Omega =2 \theta \sin (\frac{\Delta }{2})\approx \theta \Delta $ for
small $\Delta $. The probability, $p$, that a star placed at random in
the sample region falls within the great circle segment is the ratio
between the area of the segment, $\Omega $, and the area, $\Omega
_{sample}$, of the entire sample region (Figure \ref{fig: graphic})

\begin{equation}
p=\frac{\Omega }{\Omega _{sample}}=\frac{2\theta \sin (\frac{\Delta 
}{2%
})}{\Omega _{sample}}\approx \frac{\theta \Delta }{\Omega _{sample}}%
~\text{, for small }\Delta \text{.}  \label{Binomial p}
\end{equation}%

For $n$ stars randomly distributed in the sample region, the number,
$k$, of stars in the great circle segment follows a binomial
distribution with expected value $E(X)=np$ and variance
$Var(X)=np(1-p)$. The cumulative binomial distribution gives the
probability, $ p_{GC} $, of observing $k$ or more stars in the great
circle segment 
\begin{equation} 
p_{GC}=\Pr (X\geq
k)=\sum_{j=k}^{n}\binom{n}{j}p^{j}\left( 1-p\right) ^{n-j}
\label{pGCeqn} 
\end{equation} 
We evaluate this expression using the
incomplete beta function \citep{PressNumRec92}. This probability
measures the statistical significance of each observed great circle
count.

We sample the survey region by placing poles on a lattice of Galactic
longitude, $l$, and latitude, $b$, derived from a nearly uniform distribution
of antipodally symmetric points on the unit sphere using the method of
\cite{Koay11}. Note that a pole and its antipode represent the same great
circle. We then record observed counts for all segments of great circles
corresponding to these poles that lie within the sample boundaries.

To account for multiple statistical tests on the same set of data and
to avoid overstating the statistical significance of the findings, we
modify the great circle count technique of \cite{Johnston96}. Using the
Bonferroni correction \citep{Abdi07}, we adjust the probability of each
individual test, $p_{GC}$ in equation~(\ref{pGCeqn}), to account for
multiple tests by multiplying the probability by the total number of
tests of the data. The Bonferroni correction protects against false
positives but at the expense of more false negatives. The Bonferroni
correction assumes that the tests are independent and represents a
lower bound when they are not. Thus, our estimate of the overall
probability of false positives is conservative. In fact, we expect that
none of the highest confidence streams we detect in the HVS survey
sample are false positives (see \S \ref{sec: GreatCircleCounts}).

\subsection{Radial Velocity Distribution \label{sec: RVs}}

Radial velocities of stars are a potentially important probe for the
presence of star streams. The velocity distribution for halo stars is
nearly Gaussian with average line-of-sight velocity $v_{rf}=0$~km~s$^{-1}$ and velocity
dispersion $\sigma _{rf}=106$~km~s$^{-1}$ in the Galactocentric rest frame
\citep{brown10}. The radial velocity dispersion for stars in a stream
should be $\lesssim 10$~km~s$^{-1}$
\citep[e.g.,][]{Harding01,Majewski04b}. Relative to the large dispersion
of the halo, this small velocity dispersion should enable detection of
streams in velocity space.

We search for evidence of streams by comparing the distribution of radial
velocities for the stars in a great circle with the {\it observed} distribution
of radial velocities for the entire sample. If the distribution of radial
velocities on the sky is randomly drawn from this underlying distribution, the
radial velocities of stars lying in an arbitrary great circle should follow the
same distribution. Rather than assuming a Gaussian radial velocity distribution
of background halo stars {\it a priori} \citep [\emph{cf.}] []
{Harding01,Schlaufman09}, we use the observed radial velocity distribution for
the sample. In this approach, the two-sample Kolmogorov-Smirnov test provides a
nonparametric estimate of the probability, $p_{V_{rf}}$, that the radial
velocity distribution for stars within a great circle is randomly drawn from
the same distribution as the HVS survey. We compute $p_{V_{rf}}$ for great
circles with poles on the same grid we use for the great circle counts. For
each segment of a great circle that lies within the boundaries of the HVS
sample, we have $p_{GC}$ and $p_{V_{rf}}$.

We combine the great circle count and radial velocity probabilities to form a
joint test under the assumption that spatial position and radial velocity are
independent. The probability of observing a group of stars and their radial
velocities within a great circle segment of width $\Delta $ is then the product
of the individual probabilities for each of these events,
$p_{Joint}=p_{GC}\times p_{V_{rf}} $. Although position and radial velocity
may not be truly independent, the combined probability gives a measure of the
joint departure of the spatial and velocity distributions from those for a set
of points randomly distributed on the sphere and randomly drawn from the
observed radial velocity distribution.

\subsection{Limitations of the Technique \label{sec: Limits}}

There are several important observational and theoretical limitations to
our ability to detect tidal streams. Observational limitations include the
limited sky coverage of a sample and the limited depth and density of a
sample. Theoretical limitations include departure of the orbit of the
stream from a great circle and multiple orbits or wraps of the streams.
\cite{Harding01} review many of the issues involved in detecting tidal
debris in the Galactic halo.

Existing radial velocity surveys sample only a fraction of the sky.  
Consequently, we can only access great circle segments with arc lengths
$\theta < 360^\circ $. The observed magnitude limit of a radial
velocity survey defines the volume of the halo and the part of the orbit we
sample. We are potentially less sensitive to the apocenter passage than to
the pericenter passage of the orbit. More stars are stripped from the
progenitor of a star stream near pericenter, probably enhancing our ability
to detect streams and possibly leading us to overestimate the density of the
stream.

Our sensitivity depends on the width of the great circle that we choose.  
The overall density of the sample sets a lower limit on the width of great
circles we can profitably explore. As we increase the width of the great
circles we use to sample the distribution, dilution of potential streams
by background stars increases. This problem is particularly serious for
the intrinsically narrowest streams.

For computational convenience we approximate orbits as great circles.
However, the sun's offset of approximately 8 kpc from the Galactic center
means that orbits do not appear exactly as great circles on the celestial
sphere, an effect that diminishes with the size of the orbit and the
proximity of the sun to the plane of the orbit. Furthermore, the orbits
depart from great circles for a nonspherical Galactic potential.

A nonspherical Galactic potential causes precession of the orbit of the
progenitor \citep{Law10a}. Stars stripped during different passes of the
progenitor may then appear in the same observed great circle. In our
technique, one indication of this phenomenon is the appearance of multiple
peaks in the observed radial velocity distribution for the great circle
approximating the orbits. Multiple wraps may also increase the great circle
count. In addition, intersection with other streams or structures may
produce multiple peaks in the radial velocity distribution as well as an
increase in the great circle count.

% Figure 2
\begin{figure*} \begin{center} \includegraphics[width=7in]{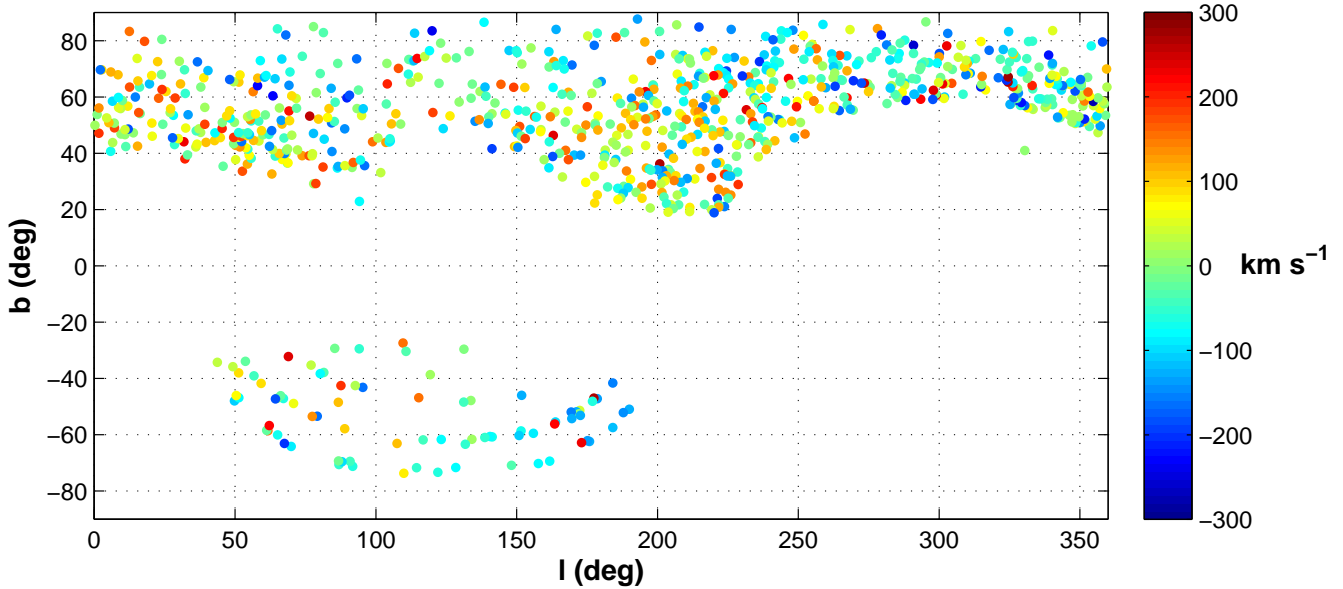} 
\caption{Galactic coordinates $ (l,b)$ of the 881 halo stars in the HVS program survey with $\left\vert v_{rf}\right\vert \protect\nolinebreak <\protect\nolinebreak 275$~km~s$^{-1}$. As indicated in the legend to the right of the main figure, color represents radial velocity in the range $-300$~km~s$^{-1}$ to $+300$~km~s$^{-1}$. \label{fig: sample} } \end{center}
\end{figure*}

\section{APPLICATION TO THE HYPERVELOCITY STAR SURVEY}
\label{sec:DataSection}

We demonstrate our technique by applying it to the spectroscopic sample of
distant halo stars from the HVS survey of \cite{brown05, brown06a, brown06b,
brown07a, brown07b, brown09a, brown09b}. The \cite{brown10} catalog of 910 late
B-type stars is 93\% complete over the magnitude range $17<g_{0}<20.5$ and
covers more than $8400$ deg$^{2}$ of the Sloan Digital Sky Survey (SDSS) Data
Release 6 (DR6) imaging footprint. These halo stars are a mix of evolved blue
stragglers and BHB stars. Because the selection is well understood, this survey
and others that cover a large area of the sky uniformly are particularly
powerful probes of the halo.

\subsection{Data \label{sec: Data}}

\cite{brown10} measured radial velocities using the Blue Channel Spectrograph
on the 6.5m MMT. They derived velocities by constructing appropriate stellar
templates and then applying standard cross-correlation routines (Kurtz \& Mink
1998). The typical radial velocity uncertainty is $\pm 12$ km s$^{-1}$.
\cite{brown10} transform heliocentric velocities $(v_{helio})$ into
Galactocentric rest frame velocities assuming a circular velocity of 220 km
s$^{-1}$ and a solar motion of $(U,V,W)=(10.0,5.2,7.2)$ km s$^{-1}$
\citep{Dehnen98}. For each star, we use this Galactocentric rest frame
velocity, $v_{rf}$.

From the 910 stars in HVS survey, we construct a sample of the 881 stars
with $\left\vert v_{rf}\right\vert <275$~km s$^{-1}$. The upper limit on
the velocity removes unbound HVSs. The 881 stars in our sample range in
heliocentric distance from 12 to 91 kpc. Figure \ref{fig: sample} shows
the distribution of the sample stars on the sky; the color of the points
indicates their radial velocities.

\subsection{Great Circle Counts\label{sec: GreatCircleCounts}}

For the HVS survey, we calculate $p_{GC}$ (equation~(\ref{pGCeqn})) for
great circles with width $\Delta = 5\degree$. The surface density of
stars in the survey constrains $\Delta$. We chose this width by
experiment. The survey is too sparse to detect streams with $\Delta < $
5\degree; adopting $\Delta \ge$ 10\degree~dilutes the signal of
streams. Other investigators use $\Delta = 5\degree$
\citep{Majewski03}, $10\degree$ \citep {Ibata01} and $12\degree$ \citep
{Ibata02b}. We distribute $2.4 \times 10^4$ antipodally symmetric
points nearly uniformly on the unit sphere \citep{Koay11},
approximately one pole per $1.7$ square degrees. Since the poles are
antipodally symmetric, this represents $1.2 \times 10^4$ unique great
circles. We consider only great circles that intersect the sample
region with arc lengths $\theta > 25\degree$. For shorter arcs, small
number statistics dominate the results. Our dense grid of sampling
poles (\S \ref{sec: Counts}) produces 9884 such great circle segments.
Many of these segments overlap.

\begin{figure} \begin{center} \includegraphics[width=3.5in]{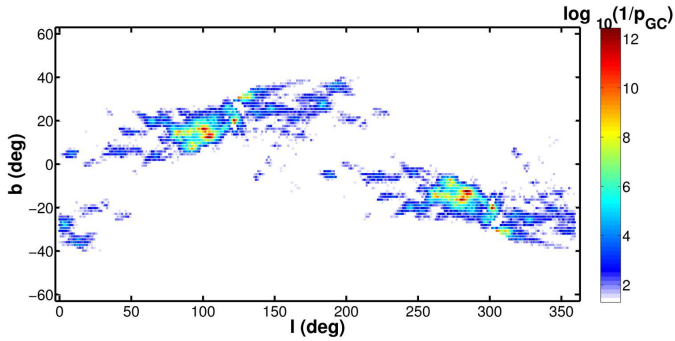} 
\caption{Significance, $\log _{10}(1/p_{GC})$, of the observed great 
circle count for pole at $(l,b)$, where $p_{GC}$\ represents the 
unadjusted probability that the observed count results from a random, 
isotropic distribution of stars, $\protect\theta \geq 25\degree$ and 
$\Delta = 5\degree$. As indicated in the legend to the right of the main 
figure, color denotes significance in the range $\log _{10}(1/p_{GC})$ = 
1.5--12.5. \label{fig: pGCraw}} \end{center} \end{figure}

Figures 3 and 4 show results for the great circle counts derived from the
unadjusted $p_{GC}$ without the Bonferroni correction (Figure \ref{fig: pGCraw})  
and the adjusted $p_{GC}$ with the Bonferroni correction (Figure \ref{fig:
pGCadj}). At the position of each pole, we encode $\log _{10}\{1/p_{GC}\} $ in
the color scale indicated in the legend. We plot only results with $p$-values of
5\% or less ($p_{GC}\leq 0.05$). The plot has an inherent symmetry as poles at
antipodes, $(l,b)$ and $(l+180\degree,-b)$, represent the same great circle.

\begin{figure} \begin{center} \includegraphics[width=3.5in]{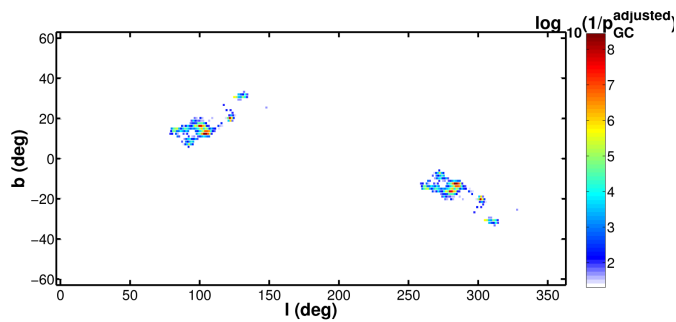} 
\caption{As in 
Figure~\ref{fig: pGCraw} for $\log _{10}(1/p_{GC}^{adjusted})$. \label{fig: pGCadj}} \end{center} \end{figure}

Comparison of these two figures illustrates the effect of accounting
for multiple statistical tests applied on the same data set. Applying
the Bonferroni correction increases the probabilities that the observed
great circle counts result from a random distribution of stars on the
sky by a factor of 9884. The values of $p_{GC}^{adjusted}$ in Figure
\ref{fig: pGCadj} range from $5.0\times 10^{-2}$ to $3.63\times
10^{-9}$.

Using the adjusted $p_{GC}$, the three most significant poles in Figure
\ref{fig: pGCadj} lie at $(103.1\degree, 12.4\degree)$, $(104.5\degree,
12.4\degree)$ and $(121.6\degree, 20.2\degree)$ with
$p_{GC}^{adjusted}=3.63\times 10^{-9}$, $4.99\times 10^{-9}$ and
$1.10\times 10^{-8}$, respectively. Features of this significance have
much less than a $1\%$ chance of appearing in a random distribution. We list 
the 25 most significant poles and identify all of them with known structures
in Table \ref{tab: ranks}. The two most significant poles coincide with
the Sagittarius stream (\S \ref{sec: JointTestResults} below).

%Figure 3

% Figure 4

\subsection{Using Radial Velocities\label{VrfResults}}

\begin{figure} \begin{center} \includegraphics[width=3.25in]{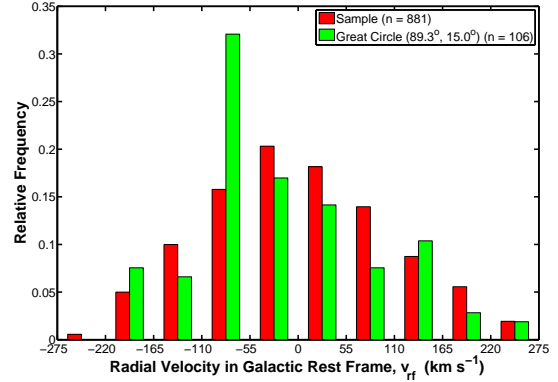}
\caption{Distribution of galactocentric radial velocities, $v_{rf}$, for sample stars in
the great circle with pole at $(l,b)=(89.3\degree,15.0\degree)$ intersecting
Sagittarius stream branch A compared with those of entire sample. The two sample Kolmogorov-Smirnov test yields a probability $p_{V_{rf}}=1.47\times 10^{-3}$ that these two distributions are drawn from the same 
distribution. \label{fig: VrfHist1} } \end{center} \end{figure}

Radial velocities provide additional information that aids in identifying
star streams, in discriminating among them, and in distinguishing them
from other structures. Figure \ref{fig: VrfHist1} illustrates the power of
comparing the velocity distribution for stars in a great circle with the
corresponding distribution for the entire HVS survey. The green histogram
compares the stellar radial velocity distribution for the great circle
with pole at $(89.3\degree, 15.0\degree)$ with the velocity distribution
for the complete survey (red histogram). For this, the most significant
structure in radial velocity in Table \ref{tab: ranks}, the two-sample
Kolmogorov-Smirnov test yields a probability of $p_{V_{rf}}=1.47\times
10^{-3}$ that the radial velocities in the great circle were drawn from
the same distribution as the entire sample.

The radial velocity distribution within this great circle has two peaks, one at
$v_{rf} \approx -80$~km~s$^{-1} $ and another at $v_{rf} \approx -190$ km
s$^{-1}$. These peaks do not appear in the distribution for the entire HVS survey
sample. The expected radial velocity dispersion within a stream should be
$\lesssim 10$ km s$^{-1}$; thus, these peaks suggest two distinct physical
components in the velocity distribution. In \S \ref{sec: JointTestResults}, 
we identify this structure in velocity with the Sagitarrius stream.

Using our HVS survey sample, we calculate the probability, $p_{V_{rf}}$,
that the radial velocities for stars in the entire sample and for stars
within a great circle come from the same distribution for all great circle
poles in our grid. Significant poles are located near the significant
great circle count pole at $(89\degree, 15\degree)$. 

% Figure 5

% Figure 6
\begin{figure} \begin{center} \includegraphics[width=3.25in]{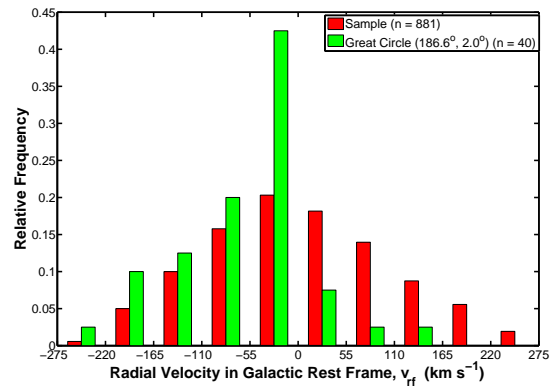}
\caption{Distribution of galactocentric radial velocities, $v_{rf}$, for sample stars in the great circle with pole at $(l,b)=(186.6\degree,2.0\degree)$
intersecting the Sagittarius stream and the Virgo overdensity compared with those of entire sample. The two sample Kolmogorov-Smirnov test yields a probability
$p_{V_{rf}}=8.31\times 10^{-5}$ that these two distributions are drawn from the same distribution. \label{fig: VrfHist2} } \end{center}
\end{figure}

The most significant pole based on radial velocity alone has
$p_{V_{rf}}=8.31\times 10^{-5}$, appears at $(186.6\degree, 2.0\degree)$,
and corresponds to a near polar great circle. Figure~\ref{fig: VrfHist2}
shows the velocity distribution of the stars in the great circle
corresponding to this pole. This detection probably results from two
separate substructures, the Sagitarrius stream and the Virgo overdensity
\citep{Newberg07} as the great circle intersects both. Some stars may belong
to the Sagitarrius stream (\S \ref{sec: SgrComparison}). Others may
belong to the Virgo overdensity, which has two peaks in its radial velocity
distribution at $v_{rf}$ = $-49$~km~s$^{-1}$ and at $v_{rf}$ =
$-171$~km~s$^{-1}$ \citep{Vivas08} that match those of stars in this great
circle.

% Figure 8
\setcounter{figure}{7}
\begin{figure*} \begin{center} \includegraphics[width=7.0in]{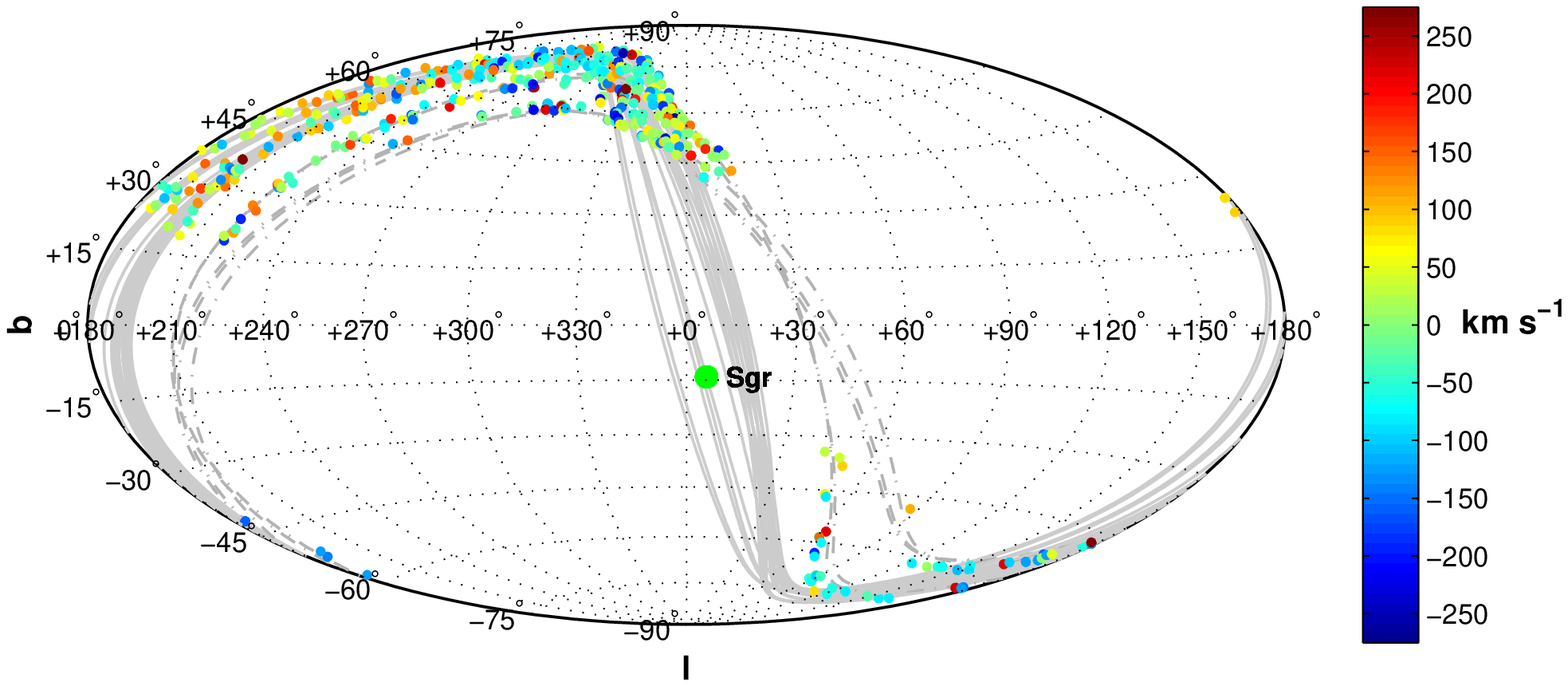}

\caption{Great circles in Galactic coordinates $(l,b)$ corresponding to 
the 25 poles with most significant \thinspace 
$p_{Joint}^{adjusted}=p_{GC}^{adjusted}\times p_{V_{rf}}$. Solid gray 
lines mark great circles affiliated with the Sagittarius stream. Dashed 
gray lines denote great circles with superpositions of the Sagittarius 
stream and other structure. The large green dot labeled Sgr marks the 
location of the Sagittarius dwarf galaxy. As indicated in the legend to 
the right of the main figure, color denotes radial velocity in km 
s$^{-1}$. \label{fig: Top25}} \end{center} 
\end{figure*}

% Figure 7
\setcounter{figure}{6}
\begin{figure} \begin{center} \includegraphics[width=3.5in]{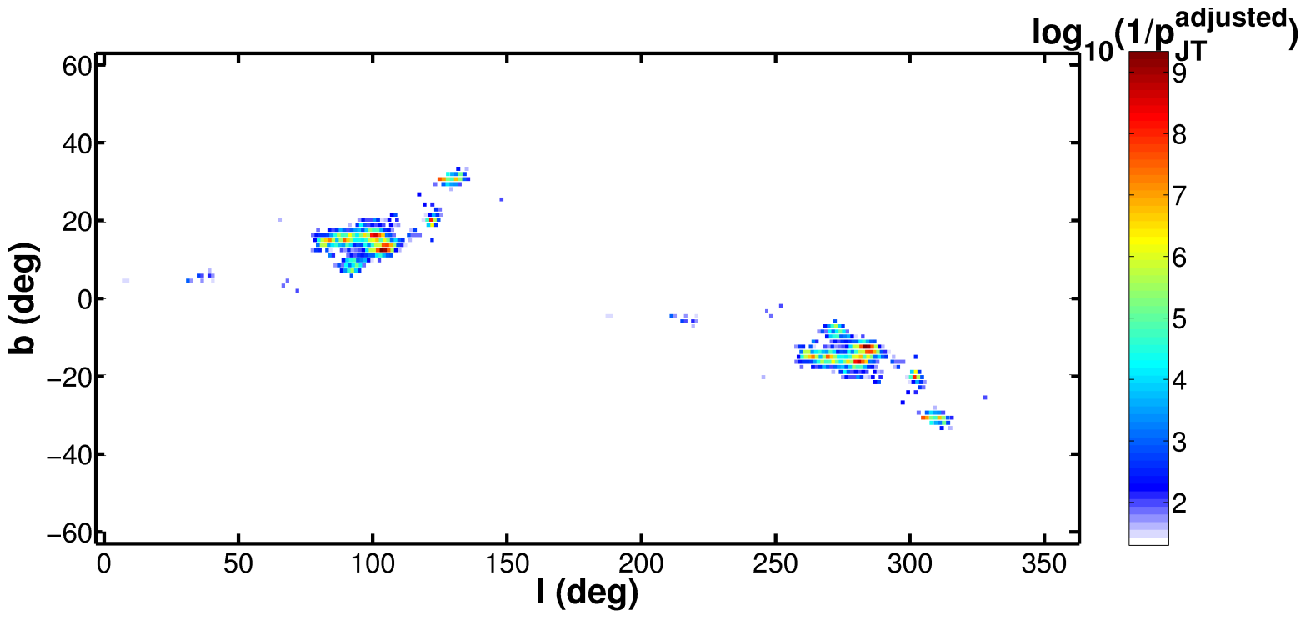}
\caption{As in Figure~\ref{fig: pGCraw} for the adjusted joint probability, $\log (1/p_{Joint}^{adjusted})$, 
where $p_{Joint}=p_{GC}^{adjusted}\times p_{V_{rf}}$.  \label{fig: pJTadj}} \end{center} \end{figure}

Next we combine both spatial and radial velocity information in a joint
statistic. Figure~\ref{fig: pJTadj} shows
$p_{Joint}^{adjusted}=p_{GC}^{adjusted}\times p_{V_{rf}} $ at the positions
of all of the poles in our grid. We plot $\log_{10}\{1/p_{Joint}^{adjusted}\}
$. Many features here are the same as in the map of significant great circle
count poles (Figure \ref{fig: pGCadj}).

We list the 25 most significant poles in $p_{Joint}^{adjusted} $ and their
associations with the Sagitarrius stream and other known structure in Table
\ref{tab:  ranks}. The two most significant poles in Figure \ref{fig: pJTadj}
and Table \ref{tab: ranks} correspond to branch~A of the Sagittarius stream
identified by \cite{Belokurov06}. These poles are also the most significant
in $p_{GC}^{adjusted}$ alone and occur at $(103.1\degree, 12.4\degree)$ and
$(104.5\degree, 12.4\degree)$ with $p_{Joint}^{adjusted}=4.63\times 10^{-10}$
and $1.09\times 10^{-9}$, respectively. Although the radial velocity is
helpful in selecting other Sagittarius poles, the values for $p_{V_{rf}}$
alone for these two poles are of low significance (Table \ref{tab: ranks}).

\subsection{Detection of the Sagittarius Stream\label{sec: JointTestResults}}

\cite{Ibata02b} analyzed M giant source counts in 26.4\% of the sky using the
2MASS Second Incremental Data Release and identified a peak corresponding to
a Sagittarius plane with pole at $(95\degree, 13\degree)$ for $\Delta $ =
$10\degree$. Using carbon star counts in great circle cells, \cite{Ibata01}
also found a peak at $(90\degree, 13\degree)$ identified with Sagittarius.
Based on their count analyses of M giants in the 2MASS survey ($\Delta =
5\degree)$, \cite{Majewski03} report poles at $(93\degree, 13\degree)$ and
$(92\degree, 12\degree)$ for the Sagittarius stream. Dividing their data set
into northern and southern Galactic hemispheres, they identify poles at
$(99\degree ,18\degree)$ and $(91.5\degree ,11.5\degree)$, respectively.

The choice of width, $\Delta $, affects the resolution of the stream. For
the $\Delta = 10\degree $ adopted in other surveys, the highly
significant points in Figure \ref{fig: pJTadj} coalesce into a single
high significance clump. Adopting $ \Delta \lesssim 5\degree$ reveals
additional structure in the stream; however, very small $\Delta \approx
1-2\degree$ results in very noisy structures and more false positives. If
the stream were simple, we could use the maximum of the probability
$p_{GC}$ as a function of the great circle width to determine its
approximate width, but this approach fails because larger widths pick up
multiple wraps of Sagittarius.

Figure \ref{fig: Top25} shows the positions and color-coded radial
velocities for sample stars in the great circles corresponding to the 25
poles with the most significant $p_{Joint}^{adjusted}$ (see Table
\ref{tab: ranks}). We also show the position of the Sagittarius dwarf
galaxy. Most of the highly significant poles are associated with the known
Sagittarius stream.

The first set of poles in Table \ref{tab: ranks} comprises branch A of the
Sagittarius stream \citep{Belokurov06}. In Figure 1 of
\citet{Belokurov06}, branch A of Sagittarius is bounded by a quadrilateral
with vertices at approximately $(\alpha ,\delta )=$ $(125\degree
,24\degree)$, $(125\degree ,17\degree)$, $(185\degree ,16\degree)$ and
$(185\degree ,7\degree)$. We identify poles corresponding to great circles
lying entirely within this region with branch A. Our two most significant
poles thus correspond to branch A. The second set of five poles is also
associated with Sagittarius branch A. The great circles for these poles
substantially overlap the bounding region of branch~A.

A third set of five poles corresponds to other parts of the Sagittarius stream.
These poles match those found in surveys of carbon stars \citep{Ibata01} and M
giant stars \citep{Majewski03}. They are also consistent with streams
identified in 2MASS \citep{Ibata02b}. Although each of these poles has a
smaller $p_{GC}^{adjusted}$ than the most significant poles in branch A of
Sagittarius, the radial velocity data provide a conclusive detection of halo
stars associated with the stream.

Despite these successes, we do not detect the Sagittarius branch~B from 
\cite{Belokurov06} in our 25 most significant poles. This failure probably 
reflects its lower surface density relative to branch~A.

The velocity distributions of stars in the great circles corresponding to
the poles in the first three sets in Table 1 are all similar to the
velocity distribution shown in Figure \ref{fig: VrfHist1} with peaks in
the velocity interval $v_{rf}\in \lbrack -110,-55] $ km s$^{-1}$.

The last set of five poles in Table 1 includes great circles that appear to
be superpositions of the Sagittarius stream with other structure. Dashed
lines in Figure \ref{fig: Top25} indicate the corresponding great circles.
The first four entries intersect the Sagittarius stream and the Virgo
overdensity \citep{Newberg07}. Thus, we suspect that both of these structures
contribute to the overdensities of stars in these four great circles.
Although the last two great circles in Table 1 appear significant in star
counts and radial velocities, their great circle arc lengths are much
shorter, $49\degree $\ and $37\degree $, than those of other great circle
segments in Table 1. Thus, we cannot robustly associate either of these great
circles with real structures in the halo.

\section{Blue Stragglers and BHB Stars in Sagittarius}
\label{sec:SgrBlueStars}

The HVS survey samples a population of blue stragglers and BHB stars that
has not been used extensively to probe the nature of the Sagittarius tidal
debris. From \S \ref{sec: JointTestResults}, the most significant
structures in the HVS survey result from Sagittarius. We estimate an upper
limit on the proportion of stars in the HVS survey sample associated with
the Sagitarrius stream directly from the sample.  By comparing the sample
of stars in our most significant streams with the $N$-body models of
\cite{Law10a}, we can (1) estimate a lower limit on the fraction of stars
in the HVS survey sample associated with the Sagittarius stream, (2)
explore the variation in stellar populations along the stream, and (3)
constrain the surface density of stars within the stream.

% Figure 9
\setcounter{figure}{8}
\begin{figure} \begin{center} \includegraphics[width=3.5in]{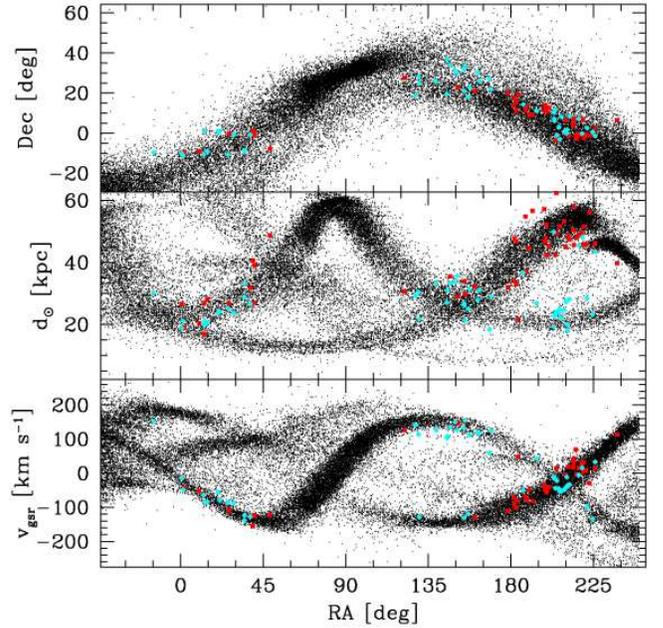}
\caption{The position, heliocentric distance, and galactocentric velocity of the \protect\cite{Law10a} Sagittarius $N$-body 
model (black dots) and the blue horizontal branch (red squares) and blue stragglers (cyan diamonds) that match the Sagittarius $N$-body model in position, velocity, and distance. \label{fig: LawModel} } \end{center} \end{figure}

\subsection{Stellar Populations and Sagittarius $N$-body Models\label{sec: SgrComparison}}

\cite{Law10a} construct and analyze a sophisticated $N$-body model of the
evolution of Sagittarius in a triaxial Milky Way potential. The model matches
all existing observational constraints on the nature of the dwarf galaxy and
the streams of debris, providing a benchmark for interpreting the apparent
debris streams we detect in the HVS survey.

We test the positions and radial velocities of HVS sample stars in the 25 most
significant great circles against the \cite{Law10a} model. Figure \ref{fig:
LawModel} shows the position, heliocentric distance, and galactocentric
velocity for the \cite{Law10a} $N$-body model (black dots). 

\cite{brown10} compute luminosity estimates for each star in the HVS survey
by matching observed colors and spectroscopy to stellar evolution tracks for
metal-poor main sequence \citep{Girardi02, Girardi04} and post-main sequence
\citep{Dotter07, Dotter08} stars. Coupled with the \cite{Law10a} model,
these luminosities enable us to discriminate between blue stragglers and BHB
stars in the HVS survey.

BHB stars in the HVS survey have a mean $M_{g}=1.15\pm 0.35$; blue stragglers
have a mean $M_{g}=2.25\pm 0.45$ \citep{brown10}. Thus, blue stragglers
should sample the nearest regions of the Sagittarius stream; BHB stars should
sample the furthest regions. Both types of stars contribute to the stream for
distances $20<d<45$ kpc.

Of the 417 stars contained in the 25 most significant great circles, 60
(14\%) match the Sagittarius $N$-body model in position, velocity, and
distance for a BHB star (red squares in Figure \ref{fig: LawModel}); 51
(12\%) match the Sagittarius $N$-body model in position, velocity, and
distance for a blue stragglers (cyan squares in Figure \ref{fig:  
LawModel}).  Seventeen stars are common to both matches. Thus 22.5\% of the
stars in our 25 most significant great circles, roughly 10\% of the total
HVS survey sample, probably belong to the Sagittarius stream. Because some
stars in the Sagittarius stream might lie in other great circles, this
estimate yields a rough lower limit to the fraction of HVS survey stars
within the Sagittarius stream.

The HVS survey stars that match the Sagittarius stream $N$-body model in the
region $0\degree <$ $RA$ $<50\degree $ are an equal mix of blue stragglers
and BHB stars. The stars that match the Sagittarius stream in the region
$130\degree <RA<180\degree $ are a 2:1 mix of blue stragglers and BHB stars
(Figure \ref{fig: LawModel}). The HVS survey fairly samples both stellar
populations within the Sagittarius stream in these regions. The observations
thus support the idea that the stellar population varies along the stream.

From color-magnitude diagrams of Sagittarius, \cite{Niederste-Ostholt10} find
equal numbers of blue stragglers and BHB stars in the region $150\degree <$
$RA<10\degree $. Detailed star count analyses, however, suggest a variation
in the number ratio of BHB stars to main sequence turn-off stars along the
stream \citep[e.g.,][]{Bell10}. Our analysis is also consistent with changes
in the stellar population along the stream. \citet{Bell10} propose that
Sagittarius once had a BHB-rich core and a BHB-poor halo. Population
variations then result from stripping different regions of the progenitor at
different times \citep[see also,][]{Martinez-Delgado04, Bellazzini06, Chou07,
Penarrubia10, Carlin11}.  Because our analysis yields matches to the position
and velocity of two stellar populations, blue stragglers and BHB stars, our
results support and extend the conclusion of \citet{Bell10} that the stellar
population in the Sagittarius dwarf varied with radius prior to disruption.

\subsection{Density of Blue Stars in the Sagittarius Stream\label{sec: BlueStarDensity}}

The \citet{Law10a} model sets a rough lower limit on the
contribution of Sagittarius to the HVS survey population. From the
observed data, we derive an approximate upper limit. To make this
estimate, we compute great circle counts for widths, $\Delta$,
between 1\degree\ and 20\degree. For each $\Delta$, we perform the
analysis outlined in \S \ref{sec:IDStreams} and identify the
great circle with the smallest $p_{Joint}$. We then compare the
number of stars within this great circle, $n_{obs}$, with the number
expected, $n_{exp}$, for a random distribution of stars on the sky.
The estimated number of stars within the Sagittarius stream is the
excess number of stars above random, $n_{Sgr}$ = $n_{obs} -
n_{exp}$. {\it A posteriori}, we verify that the locations of the
poles for this collection of great circles change little as a
function of $\Delta$ and that the excess number of stars as a
function of $\Delta$ is independent of any small change in the $(l,
b)$ of the most significant pole for an adopted $\Delta$.

\begin{figure} \begin{center} \includegraphics[width=3.25in]{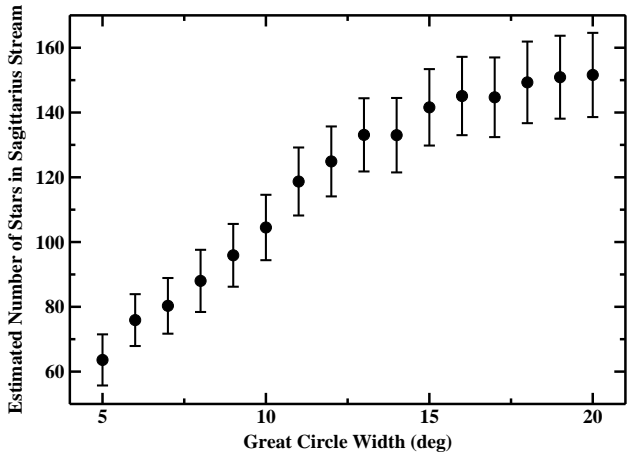}
\caption{Estimated number of stars from the HVS survey within Sagittarius as a function of great 
circle width $\Delta$.} \label{fig: SgrCounts} \end{center} \end{figure}

Figure \ref{fig: SgrCounts} shows our estimate for the number of
survey stars in the Sagittarius stream, $n_{Sgr}$, as a function of
$\Delta$. The derived $n_{Sgr}$ rises from $\sim$ 70 for $\Delta =
5\degree$ to $\sim$ 150 for $\Delta$ = 14\degree\ and then levels
off. Adopting a typical $n_{Sgr} \approx 150$ for $\Delta$ =
14\degree--20\degree, this result implies that roughly 17\% (150
stars out of 881) in the HVS survey sample belong to the Sagittarius
stream. This estimate is larger than the $\sim$ 10\% derived from
the comparison of our 25 most significant great circles with the
\cite{Law10a} model in \S \ref{sec: SgrComparison}. Combining
these results shows that Sagittarius contributes about 10\% to 17\%
of the stars in the HVS survey.

Because this approach uses the full HVS survey, it yields a better
estimate of the fraction of HVS survey stars within the Sagittarius
stream.

% Figure 10

\subsection{Stellar Surface Density of the Sagittarius Stream \label{sec: SgrSurfDen}}

To conclude this section, we compare the surface density of stars identified
in our observations of the Sagittarius stream with the predictions of the
\cite{Law10a} model. For each iteration in a Monte Carlo simulation, we
construct a sample of model stars with the same number of stars observed in
the HVS survey. This model sample consists of (i) stars drawn randomly from
the \cite{Law10a} model and (ii) stars distributed randomly and
isotropically in the HVS survey area. Based on our result in \S
\ref{sec: BlueStarDensity}, we select 17\% of this sample from the
\cite{Law10a} model. We then analyze great circle counts using the same
procedure as for the HVS survey sample with $\Delta = 5 \degree$. Repeating
this process $200$ times, we derive the median value of $p_{GC}$\ for the
most significant pole in the model Sagittarius stream.

Using equation \ref{Binomial p}, we can express the ratio of the observed
surface density to that predicted from the Monte Carlo simulations as $\gamma
=\frac{n_{obs}p_{MC}}{n_{MC}p_{obs}}$. Here, $n_{obs}$ is the number of stars
in the most significant great circle observed in the HVS survey sample,
$n_{MC}$ is the median value of the number of stars in the most significant
great circle from the Monte Carlo simulations, and $p_{obs}$ and $p_{MC}$ are,
respectively, the probabilities that a star placed randomly in the sample
region falls within the observed or simulated great circle.

This analysis shows that the \cite{Law10a} model matches our observed surface
density for stars in the Sagittarius stream. For simulations with $\Delta$ =
5\degree, we derive $\gamma$ = 1.03, with a 90\% confidence interval of
0.82--1.43. Simulations with other $\Delta$'s yield similar results.

\section{CONCLUSION}
\label{sec:Conclusion}

We construct a statistical technique to detect star streams in the Galactic
halo based on both the surface number density and the radial velocity
distribution within great circles. The addition of radial velocities to the
analysis increases our ability to identify star streams and to show that
some overdensities in number counts are superpositions of several distinct
structures in velocity space. The technique is straightforward yet powerful,
can be generalized to diverse sets of data, and quantifies the statistical
significance of tidal star streams. Detection is possible even when a
structure contributes only a few percent of the stars in a sample.

We apply the method to the HVS survey and detect the Sagittarius stream at
high significance. Comparison with theoretical models confirms that blue
stragglers and BHB stars are members of the Sagittarius stream. Great circle
counts and comparisons with theoretical models suggest that the
Sagittarius stream comprises 10\% to 17\% of the halo stars in the HVS
sample.  The ratio of blue stragglers to blue horizontal branch stars varies
along the length of the stream, with roughly equal numbers of these two
stellar types at RA = 0\degree--50\degree\ and a 2:1 mix of blue stragglers
to blue horizontal branch stars at RA = 130\degree--180\degree. Our
conclusions support previous indications of variations in stellar population
related to the original structure of the dwarf galaxy.

This technique is easily applied to other large radial velocity surveys including 
RAVE \citep{Steinmetz06,Zwitter08} and SEGUE \citep{Yanny09}. Identifying tidal 
star streams in these radial velocity surveys will ultimately improve constraints on 
the Milky Way's formation history, dark matter halo mass distribution, and the 
distributions of stellar populations within the progenitors. 

% \bigskip

% \textbf{Acknowledgements}

\acknowledgements

We thank M. Alegria, J. McAfee, A. Milone, and the rest of the MMTO staff for 
their assistance with observations obtained at the MMT Observatory.  We thank David Law 
for providing us with $N$-body simulations from his model of the Sagittarius stream. 
This project makes use of data products from the Sloan Digital Sky Survey, which is managed by the Astrophysical
Research Consortium for the Participating Institutions. This research
makes use of NASA's Astrophysics Data System Bibliographic Services. This
work was supported by the Smithsonian Institution. CK gratefully
acknowledges additional support from Pleiades Consulting Group Inc. and
dedicates this work to the memory of Dr. Natarajan Visvanathan.

\textit{Facilities:} \facility{MMT (Blue Channel Spectrograph)}\pagebreak

% \bibliographystyle{apj} \bibliography{Stream}

\begin{thebibliography}{47}
\expandafter\ifx\csname natexlab\endcsname\relax\def\natexlab#1{#1}\fi

\bibitem[{{Abadi} {et~al.}(2003){Abadi}, {Navarro}, {Steinmetz}, \&
  {Eke}}]{Abadi03b}
{Abadi}, M.~G., {Navarro}, J.~F., {Steinmetz}, M., \& {Eke}, V.~R. 2003, \apj,
  597, 21

\bibitem[{{Abdi}(2007)}]{Abdi07}
{Abdi}, H. 2007, in Encyclopedia of Measurement and Statistics, ed. N.~Salkind
  (Thousand Oaks, CA: Sage),
  \url{http://www.utdallas.edu/$\sim$herve/Abdi-Bonferroni2007-pretty.pdf}

\bibitem[{{Bell} {et~al.}(2010){Bell}, {Xue}, {Rix}, {Ruhland}, \&
  {Hogg}}]{Bell10}
{Bell}, E.~F., {Xue}, X.~X., {Rix}, H.-W., {Ruhland}, C., \& {Hogg}, D.~W.
  2010, \aj, 140, 1850

\bibitem[{{Bellazzini} {et~al.}(2006){Bellazzini}, {Newberg}, {Correnti},
  {Ferraro}, \& {Monaco}}]{Bellazzini06}
{Bellazzini}, M., {Newberg}, H.~J., {Correnti}, M., {Ferraro}, F.~R., \&
  {Monaco}, L. 2006, \aap, 457, L21

\bibitem[{{Belokurov} {et~al.}(2006)}]{Belokurov06}
{Belokurov}, V., {et~al.} 2006, \apjl, 642, L137

\bibitem[{{Brown} {et~al.}(2009{\natexlab{a}}){Brown}, {Geller}, \&
  {Kenyon}}]{brown09a}
{Brown}, W.~R., {Geller}, M.~J., \& {Kenyon}, S.~J. 2009{\natexlab{a}}, \apj,
  690, 1639

\bibitem[{{Brown} {et~al.}(2009{\natexlab{b}}){Brown}, {Geller}, {Kenyon}, \&
  {Bromley}}]{brown09b}
{Brown}, W.~R., {Geller}, M.~J., {Kenyon}, S.~J., \& {Bromley}, B.~C.
  2009{\natexlab{b}}, \apjl, 690, L69

\bibitem[{{Brown} {et~al.}(2010){Brown}, {Geller}, {Kenyon}, \&
  {Diaferio}}]{brown10}
{Brown}, W.~R., {Geller}, M.~J., {Kenyon}, S.~J., \& {Diaferio}, A. 2010, \aj,
  139, 59

\bibitem[{{Brown} {et~al.}(2005){Brown}, {Geller}, {Kenyon}, \&
  {Kurtz}}]{brown05}
{Brown}, W.~R., {Geller}, M.~J., {Kenyon}, S.~J., \& {Kurtz}, M.~J. 2005,
  \apjl, 622, L33

\bibitem[{{Brown} {et~al.}(2006{\natexlab{a}}){Brown}, {Geller}, {Kenyon}, \&
  {Kurtz}}]{brown06a}
---. 2006{\natexlab{a}}, \apjl, 640, L35

\bibitem[{{Brown} {et~al.}(2006{\natexlab{b}}){Brown}, {Geller}, {Kenyon}, \&
  {Kurtz}}]{brown06b}
---. 2006{\natexlab{b}}, \apj, 647, 303

\bibitem[{{Brown} {et~al.}(2007{\natexlab{a}}){Brown}, {Geller}, {Kenyon},
  {Kurtz}, \& {Bromley}}]{brown07a}
{Brown}, W.~R., {Geller}, M.~J., {Kenyon}, S.~J., {Kurtz}, M.~J., \& {Bromley},
  B.~C. 2007{\natexlab{a}}, \apj, 660, 311

\bibitem[{{Brown} {et~al.}(2007{\natexlab{b}}){Brown}, {Geller}, {Kenyon},
  {Kurtz}, \& {Bromley}}]{brown07b}
---. 2007{\natexlab{b}}, \apj, 671, 1708

\bibitem[{{Bullock} \& {Johnston}(2005)}]{Bullock05}
{Bullock}, J.~S., \& {Johnston}, K.~V. 2005, \apj, 635, 931

\bibitem[{{Carlin} {et~al.}(2011){Carlin}, {Majewski}, {Casetti-Dinescu},
  {Law}, {Girard}, \& {Patterson}}]{Carlin11}
{Carlin}, J.~L., {Majewski}, S.~R., {Casetti-Dinescu}, {et~al.} 2011, \apj, {accepted}

\bibitem[{{Chou} {et~al.}(2007){Chou}, {Majewski}, {Cunha}, {Smith},
  {Patterson}, {Mart{\'{\i}}nez-Delgado}, {Law}, {Crane}, {Mu{\~n}oz}, {Garcia
  L{\'o}pez}, {Geisler}, \& {Skrutskie}}]{Chou07}
{Chou}, M.-Y., {Majewski}, S.~R., {Cunha}, K., {et~al.} 2007, \apj, 670, 346

\bibitem[{{Dehnen} \& {Binney}(1998)}]{Dehnen98}
{Dehnen}, W., \& {Binney}, J.~J. 1998, \mnras, 298, 387

\bibitem[{{Dotter} {et~al.}(2007){Dotter}, {Chaboyer}, {Jevremovi{\'c}},
  {Baron}, {Ferguson}, {Sarajedini}, \& {Anderson}}]{Dotter07}
{Dotter}, A., {Chaboyer}, B., {Jevremovi{\'c}}, D., {et~al.} 2007, \aj, 134, 376

\bibitem[{{Dotter} {et~al.}(2008){Dotter}, {Chaboyer}, {Jevremovi{\'c}},
  {Kostov}, {Baron}, \& {Ferguson}}]{Dotter08}
{Dotter}, A., {Chaboyer}, B., {Jevremovi{\'c}}, D., {et~al.} 2008, \apjs, 178, 89

\bibitem[{{Font} {et~al.}(2006){Font}, { Johnston}, {Bullock}, \&
  {Robertson}}]{Font06}
{Font}, A.~S., { Johnston}, K.~V., {Bullock}, J.~S., \& {Robertson}, B.~E.
  2006, \apj, 638, 585

\bibitem[{{Girardi} {et~al.}(2002){Girardi}, {Bertelli}, {Bressan}, {Chiosi},
  {Groenewegen}, {Marigo}, {Salasnich}, \& {Weiss}}]{Girardi02}
{Girardi}, L., {Bertelli}, G., {Bressan}, {et~al.} A. 2002, \aap, 391, 195

\bibitem[{{Girardi} {et~al.}(2004){Girardi}, {Grebel}, {Odenkirchen}, \&
  {Chiosi}}]{Girardi04}
{Girardi}, L., {Grebel}, E.~K., {Odenkirchen}, M., \& {Chiosi}, C. 2004, \aap,
  422, 205

\bibitem[{{Harding} {et~al.}(2001){Harding}, {Morrison}, {Olszewski},
  {Arabadjis}, {Mateo}, {Dohm-Palmer}, {Freeman}, \& {Norris}}]{Harding01}
{Harding}, P., {Morrison}, H.~L., {Olszewski}, E.~W., {et~al.} 2001, \aj,
  122, 1397

\bibitem[{{Ibata} {et~al.}(2001){Ibata}, {Lewis}, {Irwin}, {Totten}, \&
  {Quinn}}]{Ibata01}
{Ibata}, R., {Lewis}, G.~F., {Irwin}, M., {Totten}, E., \& {Quinn}, T. 2001,
  \apj, 551, 294

\bibitem[{{Ibata} {et~al.}(1994){Ibata}, {Gilmore}, \& {Irwin}}]{Ibata94}
{Ibata}, R.~A., {Gilmore}, G., \& {Irwin}, M.~J. 1994, \nat, 370, 194

\bibitem[{{Ibata} {et~al.}(2002){Ibata}, {Lewis}, {Irwin}, \&
  {Cambr{\'e}sy}}]{Ibata02b}
{Ibata}, R.~A., {Lewis}, G.~F., {Irwin}, M.~J., \& {Cambr{\'e}sy}, L. 2002,
  \mnras, 332, 921

\bibitem[{{Johnston} {et~al.}(1996){Johnston}, {Hernquist}, \&
  {Bolte}}]{Johnston96}
{Johnston}, K.~V., {Hernquist}, L., \& {Bolte}, M. 1996, \apj, 465, 278

\bibitem[{{Johnston} {et~al.}(1999){Johnston}, {Majewski}, {Siegel}, {Reid}, \&
  {Kunkel}}]{Johnston99}
{Johnston}, K.~V., {Majewski}, S.~R., {Siegel}, M.~H., {Reid}, I.~N., \&
  {Kunkel}, W.~E. 1999, \aj, 118, 1719

\bibitem[{{Koay}(2011)}]{Koay11}
{Koay}, C.~G. 2011, Journal of Computational Science, in press

\bibitem[{{Law} \& {Majewski}(2010)}]{Law10a}
{Law}, D.~R., \& {Majewski}, S.~R. 2010, \apj, 714, 229

\bibitem[{{Lynden-Bell} \& {Lynden-Bell}(1995)}]{Lynden-Bell95}
{Lynden-Bell}, D., \& {Lynden-Bell}, R.~M. 1995, \mnras, 275, 429

\bibitem[{{Majewski} {et~al.}(2003){Majewski}, {Skrutskie}, {Weinberg}, \&
  {Ostheimer}}]{Majewski03}
{Majewski}, S.~R., {Skrutskie}, M.~F., {Weinberg}, M.~D., \& {Ostheimer}, J.~C.
  2003, \apj, 599, 1082

\bibitem[{{Majewski} {et~al.}(2004){Majewski}, {Kunkel}, {Law}, {Patterson},
  {Polak}, {Rocha-Pinto}, {Crane}, {Frinchaboy}, {Hummels}, {Johnston}, {Rhee},
  {Skrutskie}, \& {Weinberg}}]{Majewski04b}
{Majewski}, S.~R., {Ostheimer}, J.~C., {Rocha-Pinto}, H.~J., {et~al.} 2004, \aj, 128, 245

\bibitem[{{Mart{\'{\i}}nez-Delgado} {et~al.}(2004){Mart{\'{\i}}nez-Delgado},
  {G{\'o}mez-Flechoso}, {Aparicio}, \& {Carrera}}]{Martinez-Delgado04}
{Mart{\'{\i}}nez-Delgado}, D., {G{\'o}mez-Flechoso}, M.~{\'A}., {Aparicio}, A.,
  \& {Carrera}, R. 2004, \apj, 601, 242

\bibitem[{{Newberg} {et~al.}(2007){Newberg}, {Yanny}, {Cole}, {Beers}, {Re
  Fiorentin}, {Schneider}, \& {Wilhelm}}]{Newberg07}
{Newberg}, H.~J., {Yanny}, B., {Cole}, {et~al.} 2007, \apj, 668, 221

\bibitem[{{Niederste-Ostholt} {et~al.}(2010){Niederste-Ostholt}, {Belokurov},
  {Evans}, \& {Pe{\~n}arrubia}}]{Niederste-Ostholt10}
{Niederste-Ostholt}, M., {Belokurov}, V., {Evans}, N.~W., \& {Pe{\~n}arrubia},
  J. 2010, \apj, 712, 516

\bibitem[{{Pe{\~n}arrubia} {et~al.}(2010){Pe{\~n}arrubia}, {Belokurov},
  {Evans}, {Mart{\'{\i}}nez-Delgado}, {Gilmore}, {Irwin}, {Niederste-Ostholt},
  \& {Zucker}}]{Penarrubia10}
{Pe{\~n}arrubia}, J., {Belokurov}, V., {Evans}, N.~W., {et~al.} 2010, \mnras, 408, L26

\bibitem[{Press {et~al.}(1992)Press, Teukolsky, Vetterling, \&
  Flannery}]{PressNumRec92}
Press, W., Teukolsky, S., Vetterling, W., \& Flannery, B. 1992, Numerical
  Recipes in FORTRAN, 2nd edn. (Cambridge, UK: Cambridge University Press)

\bibitem[{{Ruhland} {et~al.}(2011){Ruhland}, {Bell}, {Rix}, \&
  {Xue}}]{Ruhland11}
{Ruhland}, C., {Bell}, E.~F., {Rix}, H., \& {Xue}, X. 2011, \apj, 731, 119

\bibitem[{{Schlaufman} {et~al.}(2009){Schlaufman}, {Rockosi}, {Allende Prieto},
  {Beers}, {Bizyaev}, {Brewington}, {Lee}, {Malanushenko}, {Malanushenko},
  {Oravetz}, {Pan}, {Simmons}, {Snedden}, \& {Yanny}}]{Schlaufman09}
{Schlaufman}, K.~C., {Rockosi}, C.~M., {Allende Prieto}, C., {et~al.} 2009, \apj, 703, 2177

\bibitem[{{Sharma} {et~al.}(2011){Sharma}, {Johnston}, {Majewski}, {Bullock},
  \& {Mu{\~n}oz}}]{Sharma11}
{Sharma}, S., {Johnston}, K.~V., {Majewski}, S.~R., {Bullock}, J., \&
  {Mu{\~n}oz}, R.~R. 2011, \apj, 728, 106

\bibitem[{{Starkenburg} {et~al.}(2009){Starkenburg}, {Helmi}, {Morrison},
  {Harding}, {van Woerden}, {Mateo}, {Olszewski}, {Sivarani}, {Norris},
  {Freeman}, {Shectman}, {Dohm-Palmer}, {Frey}, \& {Oravetz}}]{Starkenburg09}
{Starkenburg}, E., {Helmi}, A., {Morrison}, H.~L., {et~al.} 2009, \apj, 698, 567

\bibitem[{{Steinmetz} {et~al.}(2006){Steinmetz}, {Zwitter}, {Siebert},
  {Watson}, {Freeman}, {Munari}, {Campbell}, {Williams}, {Seabroke}, {Wyse},
  {Parker}, {Bienaym{\'e}}, {Roeser}, {Gibson}, {Gilmore}, {Grebel}, {Helmi},
  {Navarro}, {Burton}, {Cass}, {Dawe}, {Fiegert}, {Hartley}, {Russell},
  {Saunders}, {Enke}, {Bailin}, {Binney}, {Bland-Hawthorn}, {Boeche}, {Dehnen},
  {Eisenstein}, {Evans}, {Fiorucci}, {Fulbright}, {Gerhard}, {Jauregi}, {Kelz},
  {Mijovi{\'c}}, {Minchev}, {Parmentier}, {Pe{\~n}arrubia}, {Quillen}, {Read},
  {Ruchti}, {Scholz}, {Siviero}, {Smith}, {Sordo}, {Veltz}, {Vidrih}, {von
  Berlepsch}, {Boyle}, \& {Schilbach}}]{Steinmetz06}
{Steinmetz}, M., {Zwitter}, T., {Siebert}, A., {et~al.} 2006, \aj, 132, 1645

\bibitem[{{Vivas} {et~al.}(2008){Vivas}, {Jaff{\'e}}, {Zinn}, {Winnick},
  {Duffau}, \& {Mateu}}]{Vivas08}
{Vivas}, A.~K., {Jaff{\'e}}, Y.~L., {Zinn}, R., {et~al.} 2008, \aj, 136, 1645

\bibitem[{{Xue} {et~al.}(2011){Xue}, {Rix}, {Yanny}, {Beers}, {Bell}, {Zhao},
  {Bullock}, {Johnston}, {Rockosi}, {Koposov}, {Kang}, {Liu}, {Luo}, {Lee}, \&
  {Weaver}}]{Xue11}
{Xue}, X.-X., {Rix}, H.-W., {Yanny}, B., {et~al.} 2011, \apj, 738, 79

\bibitem[{{Yanny} {et~al.}(2009)}]{Yanny09}
{Yanny}, B., {Newberg}, H.~J., {Johnson}, J.~A., {et~al.} 2009, \apj, 700, 1282

\bibitem[{{Zwitter} {et~al.}(2008){Zwitter}, {Siebert}, {Munari}, {Freeman},
  {Siviero}, {Watson}, {Fulbright}, {Wyse}, {Campbell}, {Seabroke}, {Williams},
  {Steinmetz}, {Bienaym{\'e}}, {Gilmore}, {Grebel}, {Helmi}, {Navarro},
  {Anguiano}, {Boeche}, {Burton}, {Cass}, {Dawe}, {Fiegert}, {Hartley},
  {Russell}, {Veltz}, {Bailin}, {Binney}, {Bland-Hawthorn}, {Brown}, {Dehnen},
  {Evans}, {Re Fiorentin}, {Fiorucci}, {Gerhard}, {Gibson}, {Kelz}, {Kujken},
  {Matijevi{\v c}}, {Minchev}, {Parker}, {Pe{\~n}arrubia}, {Quillen}, {Read},
  {Reid}, {Roeser}, {Ruchti}, {Scholz}, {Smith}, {Sordo}, {Tolstoi},
  {Tomasella}, {Vidrih}, \& {Wylie-de Boer}}]{Zwitter08}
{Zwitter}, T., {Siebert}, A., {Munari}, U., {et~al.} 2008, \aj, 136, 421

\end{thebibliography}

% \clearpage

\begin{deluxetable*}{lccccccc}
\tablecolumns{8}
\tablewidth{0pc}
\tabletypesize{\scriptsize}
\tablecaption{The 25 Most Significant Poles Ranked by $p_{Joint}^{adjusted}$}
\tablehead{
\colhead{$l$ (deg)} &
\colhead{$b$ (deg)} &
\colhead{Rank} &
\colhead{$p_{GC}$} &
\colhead{$p_{V_{rf}}$} &
\colhead{$p_{Joint}$} &
\colhead{$p_{GC}^{adjusted}$} &
\colhead{$p_{Joint}^{adjusted}$}}
\startdata
\cutinhead{Sagittarius Stream -- Branch A}
101.0  &  16.3  &   3  &  $2.29 \times 10^{-12}$  &  $1.11 \times 10^{-01}$  &  $2.56 \times 10^{-13}$ &   $2.27 \times 10^{-08}$ & $2.53 \times 10^{-09}$ \\
101.6  &  15.0  &  16  &  $9.44 \times 10^{-11}$  &  $1.62 \times 10^{-01}$  &  $1.53 \times 10^{-11}$ &   $9.33 \times 10^{-07}$ & $1.51 \times 10^{-07}$ \\
102.4  &  16.3  &  11  &  $6.61 \times 10^{-11}$  &  $4.77 \times 10^{-02}$  &  $3.16 \times 10^{-12}$ &   $6.54 \times 10^{-07}$ & $3.12 \times 10^{-08}$ \\
103.1  &  12.4  &   1  &  $3.67 \times 10^{-13}$  &  $1.28 \times 10^{-01}$  &  $4.69 \times 10^{-14}$ &   $3.63 \times 10^{-09}$ & $4.63 \times 10^{-10}$ \\
103.7  &  16.3  &  19  &  $5.05 \times 10^{-10}$  &  $4.92 \times 10^{-02}$  &  $2.49 \times 10^{-11}$ &   $4.99 \times 10^{-06}$ & $2.46 \times 10^{-07}$ \\
103.9  &  13.7  &   9  &  $1.35 \times 10^{-11}$  &  $1.89 \times 10^{-01}$  &  $2.56 \times 10^{-12}$ &   $1.34 \times 10^{-07}$ & $2.53 \times 10^{-08}$ \\
104.3  &  15.0  &  24  &  $6.05 \times 10^{-10}$  &  $1.17 \times 10^{-01}$  &  $7.08 \times 10^{-11}$ &   $5.98 \times 10^{-06}$ & $7.00 \times 10^{-07}$ \\
104.5  &  12.4  &   2  &  $5.05 \times 10^{-13}$  &  $2.18 \times 10^{-01}$  &  $1.10 \times 10^{-13}$ &   $4.99 \times 10^{-09}$ & $1.09 \times 10^{-09}$ \\
105.3  &  13.7  &   6  &  $1.47 \times 10^{-11}$  &  $1.12 \times 10^{-01}$  &  $1.65 \times 10^{-12}$ &   $1.45 \times 10^{-07}$ & $1.63 \times 10^{-08}$ \\
106.6  &  13.7  &  15  &  $4.69 \times 10^{-11}$  &  $2.85 \times 10^{-01}$  &  $1.34 \times 10^{-11}$ &   $4.63 \times 10^{-07}$ & $1.32 \times 10^{-07}$ \\
\cutinhead{Associated with Sagittarius Stream -- Branch A}
 96.9  &  16.3  &  21  &  $2.55 \times 10^{-10}$  &  $1.19 \times 10^{-01}$  &  $3.04 \times 10^{-11}$ &   $2.52 \times 10^{-06}$ & $3.01 \times 10^{-07}$ \\
 99.6  &  16.3  &   4  &  $2.18 \times 10^{-12}$  &  $3.17 \times 10^{-01}$  &  $6.91 \times 10^{-13}$ &   $2.15 \times 10^{-08}$ & $6.83 \times 10^{-09}$ \\
100.2  &  15.0  &  25  &  $7.93 \times 10^{-10}$  &  $9.01 \times 10^{-02}$  &  $7.15 \times 10^{-11}$ &   $7.84 \times 10^{-06}$ & $7.07 \times 10^{-07}$ \\
101.8  &  12.4  &   8  &  $3.13 \times 10^{-11}$  &  $7.12 \times 10^{-02}$  &  $2.23 \times 10^{-12}$ &   $3.09 \times 10^{-07}$ & $2.20 \times 10^{-08}$ \\
105.8  &  12.4  &  10  &  $1.02 \times 10^{-11}$  &  $2.70 \times 10^{-01}$  &  $2.76 \times 10^{-12}$ &   $1.01 \times 10^{-07}$ & $2.73 \times 10^{-08}$ \\
\cutinhead{Associated with Sagittarius Stream}
 82.5  &  15.0  &  20  &  $9.98 \times 10^{-10}$  &  $2.61 \times 10^{-02}$  &  $2.60 \times 10^{-11}$  &  $9.86 \times 10^{-06}$ & $2.57 \times 10^{-07}$ \\
 83.9  &  15.0  &  13  &  $4.20 \times 10^{-10}$  &  $1.35 \times 10^{-02}$  &  $5.67 \times 10^{-12}$  &  $4.15 \times 10^{-06}$ & $5.60 \times 10^{-08}$ \\
 89.3  &  15.0  &  12  &  $2.43 \times 10^{-09}$  &  $1.47 \times 10^{-03}$  &  $3.56 \times 10^{-12}$  &  $2.40 \times 10^{-05}$ & $3.52 \times 10^{-08}$ \\
 90.7  &  15.0  &  18  &  $7.70 \times 10^{-09}$  &  $3.08 \times 10^{-03}$  &  $2.37 \times 10^{-11}$  &  $7.61 \times 10^{-05}$ & $2.34 \times 10^{-07}$ \\
 91.4  &  16.3  &  22  &  $1.99 \times 10^{-08}$  &  $2.79 \times 10^{-03}$  &  $5.55 \times 10^{-11}$  &  $1.96 \times 10^{-04}$ & $5.48 \times 10^{-07}$ \\
\cutinhead{Superpositions with Sagittarius}
121.6  &  20.2  &   5  &  $1.12 \times 10^{-12}$  &  $9.97 \times 10^{-01}$  &  $1.11 \times 10^{-12}$  &  $1.10 \times 10^{-08}$ & $1.10 \times 10^{-08}$ \\
122.1  &  18.9  &  23  &  $1.29 \times 10^{-10}$  &  $4.88 \times 10^{-01}$  &  $6.31 \times 10^{-11}$  &  $1.28 \times 10^{-06}$ & $6.24 \times 10^{-07}$ \\
124.9  &  30.7  &   7  &  $2.14 \times 10^{-10}$  &  $1.01 \times 10^{-02}$  &  $2.16 \times 10^{-12}$  &  $2.11 \times 10^{-06}$ & $2.14 \times 10^{-08}$ \\
126.4  &  30.7  &  14  &  $5.21 \times 10^{-10}$  &  $1.75 \times 10^{-02}$  &  $9.12 \times 10^{-12}$  &  $5.15 \times 10^{-06}$ & $9.01 \times 10^{-08}$ \\
131.0  &  30.7  &  17  &  $1.21 \times 10^{-09}$  &  $1.67 \times 10^{-02}$  &  $2.01 \times 10^{-11}$  &  $1.19 \times 10^{-05}$ & $1.99 \times 10^{-07}$ \\
\enddata
\label{tab: ranks}
\end{deluxetable*}

\end{document}